%% file: EUSIPCO21.tex
\documentclass[conference]{IEEEtran}
\IEEEoverridecommandlockouts
\usepackage{amsmath,graphicx}
\usepackage{pdfpages}
\usepackage{multicol}
\usepackage{xcolor,colortbl}
\usepackage{mleftright}
\usepackage{tikz}
\usepackage{comment}
\input{math-macros}

\usepackage[linesnumbered,ruled,vlined]{algorithm2e}
\SetKwInput{KwInput}{Input}                
\SetKwInput{KwOutput}{Output}  
\usepackage{cite}
\usepackage{amsmath,amssymb,amsfonts}
\usepackage{algorithmic}
\usepackage{graphicx}
\usepackage{textcomp}
\usepackage{xcolor}
\def\BibTeX{{\rm B\kern-.05em{\sc i\kern-.025em b}\kern-.08em
    T\kern-.1667em\lower.7ex\hbox{E}\kern-.125emX}}
\usepackage{hyperref}
\begin{document}

\title{Audio Inpainting via $\ell_1$-Minimization\\ and Dictionary Learning\\
\thanks{This work was supported by the Vienna Science and Technology Fund (WWTF) project MA16-53 (INSIGHT) and the Austrian Science Fund (FWF) project I 3067-N30 (MERLIN) and Y 551-N13 (FLAME).
\emph{This is the Author's Accepted Manuscript version of work presented at EUSIPCO21. It is licensed under the terms of the \href{https://creativecommons.org/licenses/by/4.0/}{Creative Commons Attribution 4.0 International License}, which
permits unrestricted use, distribution, and reproduction in any medium, provided the original author and source are credited. The published version is available at:} \url{https://ieeexplore.ieee.org/abstract/document/9616132}}
}

\author{

\IEEEauthorblockN{ Shristi Rajbamshi}
\IEEEauthorblockA{\textit{Acoustic Research Institute} \\
\textit{Austrian Academy of Sciences}\\
Vienna, Austria \\
shristi.rajbamshi@oeaw.ac.at}
\and
\IEEEauthorblockN{ Georg Tauböck}
\IEEEauthorblockA{\textit{Acoustic Research Institute} \\
\textit{Austrian Academy of Sciences}\\
Vienna, Austria \\
georg.tauboeck@oeaw.ac.at}
\and
\IEEEauthorblockN{ Nicki Holighaus}
\IEEEauthorblockA{\textit{Acoustic Research Institute} \\
\textit{Austrian Academy of Sciences}\\
Vienna, Austria \\
nicki.holighaus@oeaw.ac.at}
\and
\IEEEauthorblockN{Peter Balazs}
\IEEEauthorblockA{\textit{Acoustic Research Institute} \\
\textit{Austrian Academy of Sciences}\\
Vienna, Austria \\
peter.balazs@oeaw.ac.at}
}

\maketitle
\begin{abstract}
Audio inpainting refers to signal processing techniques that aim at restoring missing or corrupted consecutive samples in audio signals. Prior works have shown that $\ell_1$- minimization with appropriate weighting is capable of solving audio inpainting problems, both for the analysis and the synthesis models. These models assume that audio signals are sparse with respect to some redundant dictionary and exploit that sparsity for inpainting purposes. Remaining within the sparsity framework, we utilize dictionary learning to further increase the sparsity and combine it with weighted $\ell_1$-minimization adapted for audio inpainting to compensate for the loss of energy within the gap after restoration. Our experiments demonstrate that our approach is superior in terms of signal-to-distortion ratio (SDR) and objective difference grade (ODG) compared with its original counterpart.
\end{abstract}

\section{Introduction}
\label{sec:intro}

Audio signals are often prone to distortions resulting in modification or loss of information at certain sections. These localized distortions occur mostly during recording, transmission or storage and may span from several milliseconds to few seconds. They may be caused by, e.g., impulsive noise, packet loss, or scratches on the storage device. In the field of audio signal processing, the task of recovering such a lost section aka \emph{gap} while ensuring that the audio artifact is almost imperceptible is known as \emph{audio inpainting} \cite{adler12}. We note that this recovery has also been referred to in the literature as audio interpolation/extrapolation, waveform substitution, and, more generally, for arbitrary (i.e., not necessarily audio) signals as \emph{error concealment}.

Over the years, several audio inpainting methods have been proposed, based on various approaches like auto regression by Janssen \cite{AR_audio86}, Bayesian estimation \cite{godsill1995bayesian}, convex optimization \cite{tauboeck2020audio,mokry2020audio, mokry2019introducing}, sinusoidal modeling \cite{lagrange2005long}, similarity graphs \cite{perraudin2018inpainting}, and deep neural networks \cite{marafioti2019audio,Marafioti2019,Marafioti2021}. Whereas auto regression, Bayesian estimation and convex optimization are applicable to short gaps (less than 100 ms), sinusoidal modeling, similarity graphs and deep neural networks work well for longer gaps (more than 100 ms).

In this paper, we focus on audio inpainting approaches which are motivated by the observation that real-world audio signals may be sparsely represented in some overcomplete dictionary \cite{adler12,tauboeck2020audio,mokry2020audio}. The cited methods exploit sparsity to fill the gap by employing optimization techniques while utilizing the information present around the gap. There are many different possibilities to formulate an audio inpainting problem as a sparsity-based optimization task. For example, by greedy heuristics such as OMP \cite{adler2011audio} and convex optimization via $\ell_1$-minimization \cite{mokry2020audio}. 


\subsection{Notation}\label{ssec_not}
Roman letters $\matA,\matB,\ldots$, $\veca,\vecb,\ldots$, and $a,b,\ldots$
designate matrices, vectors, and scalars, respectively. Sets will be denoted by calligraphic letters such as $\setS$. We use the notation $\matA_{\setS}$ to indicate the column submatrix of $\matA$ consisting of the columns indexed by $\setS$. $A_{k l}$ denotes the element  in the $k$th row and $l$th column of $\matA$.
The superscripts $\tp{}$ and $\herm{}$ represent transposition and Hermitian transposition, respectively. 
$\matI_N$ denotes the $N \times N$ identity matrix. We write for the indicator function $\chi_{_\setS}(\vecx)$, where $$\chi_{_\setS}(\vecx) = \left\{ \begin{array}{c c} 0 & \vecx \in \setS \\ + \infty & \text{otherwise.} \end{array} \right.$$
For a vector $\vecu=\tp{(u_0,u_1, \cdots, u_{N-1})}$
 its support (i.e., indices of non-zero entries) is denoted by $\text{supp}(\vecu)$. Its $\ell_0$-``norm'' and the $\ell_1$-norm are given as $\vecnorm{\vecu}_0
 =\card(\text{supp}(\vecu))$ (i.e., the cardinality of the support) and  $\vecnorm{\vecu}_1=\sum_n |u_n|$, respectively.
Finally, we denote by $\|\matX\|_{1,1} = \sum_{m,n} |X_{m,n}|$ the entry-wise $\ell_1$-norm of a matrix $\matX$.

\section{Motivation and Contribution}

 This work is inspired by two model-based optimization approaches for audio inpainting: \emph{SParse Audio INpainter (SPAIN) with dictionary learning}  \cite{tauboeck2020audio}, see also \cite{tauboeck2020sparse}, and \emph{ (weighted) $\ell_1$-minimization based audio inpainting}  \cite{mokry2020audio}. Note that both  methods use the Gabor transform (see Section \ref{sec: Gabor}) to obtain a sparse representation of the original audio signal. 
 
The optimization based inpainting methods are quite efficient in filling short gaps, but as the gaps tend to get longer, they suffer from energy loss within the gap after restoration. As a means to tackle this problem, remaining in the framework of sparse audio inpainting, in \cite{mokry2020audio}, various weighting methods were combined with $\ell_1$-minimization, whereas in \cite{tauboeck2020audio}, a dictionary learning method was proposed and combined with a modified version of the SPAIN algorithm, i.e., SPAIN-MOD \cite{tauboeck2020audio}. 
This combination of dictionary learning and SPAIN-MOD will be referred to as SPAIN-LEARNED henceforth. SPAIN-LEARNED deforms the underlying Gabor dictionary such that the sparsity of the analysis coefficients of the resulting dictionary is further enhanced. A short description on how the Gabor dictionary is deformed, i.e, how the dictionary is learned from Gabor coefficients is given in \ref{ssec_dict}. 
SPAIN-LEARNED exhibits excellent reconstruction performance compared to other techniques like (plain) SPAIN \cite{mokry2019introducing} and weighted $\ell_1$-methods as presented in  \cite{mokry2020audio}.

Motivated by these observations, we propose to slightly modify a selected weighted  $\ell_1$-method from \cite{mokry2020audio} as well as to incorporate the dictionary learning technique presented in \cite{tauboeck2020audio}. As we will demonstrate, this modified weighted $\ell_1$-method exhibits significantly improved reconstruction performance compared with the original setting and achieves comparable results as the (currently) state-of-the-art SPAIN-LEARNED algorithm.

\section{Gabor Systems}
\label{sec: Gabor}
In this work, we will exploit time-frequency (TF) sparsity, specifically with respect to Gabor dictionaries. In a finite discrete setting, the TF coefficients are obtained by applying the discrete Gabor transform (DGT) \cite{LTFAT1} to the input audio signal. Note that the DGT is also commonly known as short-time Fourier transform (STFT). 

For a window function $\vecg \in \reals^{L}$ and time and frequency hop sizes $a$ and $b$, the collection of TF modulated versions of $\vecg$, is called a \emph{Gabor system}, i.e.,
$$ \mathcal G = \{\vecg[\cdot-na]e^{i2\pi mb ~ \cdot/L }\}_{m,n} = \{\vecg_p\}_{p=0, 1, \ldots, NM-1}.$$
 Here, $m = 0, 1, \ldots, M-1$ is the frequency modulation index, $n = 0, 1, \ldots, N-1$ is the time index, and 
 \begin{equation}
 \label{stacking}
 p := p_{N,M} (m,n) = n + mN
\end{equation}is the combined time-frequency index. The matrix $\matG_{\text{S}}\in\complexset^{L\times NM}$, defined by setting its $p$-th column equal to $\vecg_{p}$, is denoted as \emph{synthesis operator} of the Gabor system $\mathcal G$. With the \emph{analysis operator} $\matG_{\text{A}}=\herm{\matG_{\text{S}}}$, the Gabor coefficients of a signal $\vecx\in\reals^L$ with respect to $\mathcal G$ are given by 
 \begin{equation*}
    c_p = (\matG_{\text{A}} \vecx)_p = \herm{\vecg_p} \cdot \vecx,
\end{equation*}
 where $\vecc=\tp{(c_0,c_1, \cdots, c_{NM-1})}$is called the coefficient vector of $\vecx$ and contains $P = MN$ elements.
 An appropriate choice of $\vecg$ as well as parameters $a$ and $M$ will ensure that $\matG_{\text{A}}$ has a bounded left-inverse, or equivalently, that $\mathcal{G}$ is a \emph{frame} for $\complexset^L$ \cite{christensen2008frames}. In other words, it allows perfect reconstruction from the redundant coefficients $\vecc$. In that case, the composition $\matF=\matG_{\text{S}}\matG_{\text{A}}$, called the \emph{frame operator}, is invertible and perfect reconstruction is achieved by the Gabor system generated from the {\em dual window} $\matF^{-1} \vecg$. Most DGTs used in practice form a \emph{tight} frame, i.e., $\matF = A\matI_L$, for some $A>0$, such that the dual system is generated by $(1/A)\vecg$. In this contribution, we restrict to such tight frames in order to remain consistent with the reference paper \cite{mokry2020audio} for comparison later on.
 
 Since we assume that $\vecx,\vecg\in\reals^{L}$, it is easy to see that we have 
\begin{align}\label{conj_symm}
  \vecc_{n + mN} = \overline{\vecc_{n + (M-m)N}}, \text{ for all } m\in [1,\lceil M/2\rceil -1].
 \end{align}
 Hence, it is sufficient to compute only the first $P'=N M'$ values of $\vecc$, where $M' = \lfloor M/2\rfloor +1$, see also \cite{LTFAT1,ltfatnote030}. Similarly, the analysis and synthesis operators can, for most purposes, be truncated at the $NM'$-th row and column, respectively. We will denote these truncated matrices by $\tilde{\matG}_{\text{A}}$ and $\tilde{\matG}_{\text{S}}$ and will refer to the corresponding transform as real DGT.

\section{Problem Formulation}
For  an audio signal $\vecx \in \reals^{L}$, let the indices of the \emph{reliable} and \emph{unreliable}  samples (= signal vs. gap resp.) be known. The goal of audio inpainting is to recover a
signal, $\vecy \in \reals^{L}$ from $\vecx \in \reals^{L}$ such that the gap is filled (in a `meaningful' way) and is also consistent with the reliable part of ${\vecx}$ . The (convex) set $\setS_{\vecx}$ consisting of all such possible signals can expressed as,
\begin{align*}
	&\setS_{\vecx} \define \left\{\vecy \in \reals^{L} : \matM_{\text{R}} \vecy  =  \matM_{\text{R}} \vecx\right\},
\end{align*}
where $\matM_{\text{R}}: \reals^{L} \rightarrow \reals^{L}$ is a binary masking operator that sets unreliable samples to zero and keeps only the reliable samples.

A formulation of the sparse audio inpainting problem for an analysis model is,
\begin{equation}
 \label{eq:el_0a}
\argmin_{\vecz } \|\tilde{\matG}_{\text{A}}\vecz\|_{0} \quad \text{s.t.} \quad \vecz \in {\setS_{\vecx}}.
\end{equation}

  The $\ell_0$-minimization problem described in \eqref{eq:el_0a} is NP hard. Hence, instead, one normally opts for its convex relaxation, i.e., $\ell_1$-minimization.

\section{ Audio inpainting with weighted $\ell_1$-minimization } \label{sec: l1 audio inpainting}
Introducing a \emph{weighting vector} 
\begin{align}
\label{eq:weighting_vector}
\tilde{\vecw} = \tp{(\tilde{w}_0,\ldots,\tilde{w}_{NM'-1})} \in \reals^{N M' },
\end{align}
see also \cite{xxljpa1}, a sparsity-based formulation of audio inpainting with $\ell_1$-minimization is given as,
\begin{align}
\label{eq:well_1}
\argmin_{\vecz } \|\tilde{\vecw}\odot\tilde{\matG}_{\text{A}}\vecz\|_{1} \quad \text{s.t.} \quad \vecz \in {\setS_{\vecx}}.
\end{align}
where the symbol $\odot$ denotes the element-wise product. 

We will discuss the choice of weighting vector $\tilde{\vecw}$ in the next section. 
Instead of \eqref{eq:well_1}, we can consider the unconstrained problem 
\begin{align}\label{eq:well_1cp}
&\argmin_{\vecz } \|\tilde{\vecw}\odot\tilde{\matG}_{\text{A}}\vecz\|_{1} +\chi_{_{\setS_{\vecx}}}(\vecz), 
\end{align}
which can be solved using a proximal algorithm \cite{combettes2011proximal, condat2014generic}. Proximal methods are efficient in iteratively solving complex large-scale convex minimization problems by solving a series of more, but smaller and simpler convex optimization problems.

Note that in \eqref{eq:well_1cp}, if the real DGT, $\tilde{\matG}_{\text{A}}$, and corresponding (short) weighting vector $\tilde{\vecw}\in \reals^{NM'}$ were replaced by the full DGT, ${\matG}_{\text{A}}$, and another (long) weighting vector, $\vecw=\tp{(w_0,\ldots, w_{MN-1})}\in \reals^{N M }$,  i.e.,
\begin{align}\label{eq:well_1cp_orig}
&\argmin_{\vecz } \|\vecw\odot{\matG}_{\text{A}}\vecz\|_{1} +\chi_{_{\setS_{\vecx}}}(\vecz), 
\end{align}
then \eqref{eq:well_1cp_orig} would be identical to the weighted $\ell_1$-minimization problem  solved 
in \cite{mokry2020audio} by means of the Chambolle-Pock (CP) \cite{chambolle2011first}. 
However, in order to be compatible with our dictionary learning framework according to \cite{tauboeck2020audio}, see also Subsection \ref{ssec_dict}, which is explicitly tailored to the real-valuedness of audio signals, we will consider \eqref{eq:well_1cp} instead.

\section{Choice of Weighting Vector}

Larger coefficients are often penalized more heavily than smaller coefficients in $\ell_1$-based minimization problems. In audio inpainting, this behaviour causes an energy loss within the gap in  the  restored  signal\cite{hastie2015statistical, candes2008enhancing}. 

Besides this universal effect of $\ell_1$-minimization, the choice of the overcomplete dictionary, i.e., in our case, the Gabor transform also contributes to the energy drop phenomenon. This is because the coefficients  corresponding to the window that overlap with the gap carry significantly less information about the reliable signal in comparison to the coefficients obtained from the reliable parts. In \cite{mokry2020audio}, in order to compensate for this loss, the Gabor atoms are weighted. Smaller weights are assigned to atoms carrying more unreliable information so that they contribute less to the objective function, thereby getting less penalized during minimization. 

There are several approaches to obtain a weighting vector $\vecw$, such as support-based, $\ell_1$-norm-based, $\ell_2$-norm-based, and energy-based approaches, see  \cite{mokry2020audio}. In this work, we will only consider \emph{energy-based} weighting, as in pilot tests we found this choice to be the most promising.
 If ${\vecg}_p$ is a Gabor atom  and $\matM_{\text{R}}{\vecg_p}$ is its part corresponding to the reliable part of the signal, then {energy-based} weighting according to \cite{mokry2020audio} assigns the weights $\vecw = \tp{({w}_0,\ldots,{w}_{P-1})}$ for \eqref{eq:well_1cp_orig} with
\begin{equation}\label{wheights}
    w_{p} = \frac{\|\matM_\text{R}\vecg_{p}\|_{2}^2}{\|{\vecg_{p}}\|_{2}^2}.
\end{equation}

Similarly, for \eqref{eq:well_1cp}, the weights introduced in \eqref{eq:weighting_vector}, $\tilde{\vecw} = \tp{(\tilde{w}_0,\ldots,\tilde{w}_{P'-1})}$ are simply given by \eqref{wheights} with  $\tilde{w}_{p}=w_{p}$.

\section{Learned Weighted $\ell_1$-Minimization}
\subsection{Dictionary Learning}
\label{ssec_dict}
In \cite{tauboeck2020audio}, we increased TF sparsity by means of a sparsifying dictionary, obtained by a suitable deformation of the Gabor dictionary. The applied deformation matrix was learned from the TF coefficients in the neighborhood $\setN$ of the gap, via a basis optimization method developed in \cite{7080744,GT_jstsp10}.

Succinctly, the dictionary learning technique in \cite{tauboeck2020audio} solves
\begin{equation}
    \hat{\matD} = \argmin_{\matD\in\setD} \left\|\matD \matX_\setN\right\|_{1,1},\vspace*{0mm}
\end{equation}
where $\setN\subset\{0,\ldots,N-1\}$ is a temporal neighborhood of the gap and $\matX_\setN\in \complexset^{M'\times |\setN|}$ denotes the Gabor analysis coefficients in that neighborhood, arranged in matrix form, i.e., $(\matX_\setN)_{m+1,n+1} = (\tilde{\matG}_{\text{A}} \vecx)_{n+mN}$, for $n\in\setN$ and $m\in\{0,\ldots,M'-1\}$. Further, $\setD \in \complexset^{M' \times M'}$ represents a set of unitary deformation matrices with a special structure taking into account the conjugate-symmetry \eqref{conj_symm} underlying the real DGT (see \cite{tauboeck2020audio} for a detailed description). 

\subsection{Weighted $\ell_1$-Minimization with Learned Dictionary}
Utilizing conjugate-symmetry  and including the learned dictionary, we can modify the optimization problems defined in \eqref{eq:well_1} and \eqref{eq:well_1cp}. To this end, we define the following\footnote{Note that this special structure is required to cope with the matrix arrangement/stacking according to \eqref{stacking}.} block matrix 
\begin{align*}
    \hat{\matD}_{\text{block}} := \left[\begin{array}[]{lcr}
	    \hat{D}_{1,1}\matI_N & \cdots &\hat{D}_{1 ,M'}\matI_N\\
	    \vdots &\ddots & \vdots\\
	    \hat{D}_{M', 1}\matI_N & \cdots &\hat{D}_{M', M'}\matI_N
\end{array}\right]
\end{align*}
and replace \eqref{eq:well_1} and \eqref{eq:well_1cp} by the following minimization problems, respectively,
\begin{align}
 &\nonumber  \! \! \! \! \argmin_{\vecz } \|\vecw_{\text{L}}\odot\hat{\matD}_{\text{block}}\tilde{\matG}_{\text{A}}\vecz \|_{1} \quad \! \!\text{s.t} \! \!\quad \vecz \in \setS_\vecx\\
 &\label{eq:mod weighted l1 uncons} \! \! \! \!\argmin_{\vecz } \|\vecw_{\text{L}}\odot\hat{\matD}_{\text{block}}\tilde{\matG}_{\text{A}}\vecz\|_{1} +\chi_{\setS_{\vecx}}(\vecz),
\end{align}
where $\vecw_{\text{L}}\in \mathbb{R}_+^{P'}$ denotes a weighting vector  that depends on the learned dictionary.

The elements of the deformed Gabor dictionary can be expressed as $\hat{\vecg}_{p}= (\tilde{\matG}_{\text{S}}\herm{\hat{\matD}_{\text{block}}})_{p+1}$,
$p\in \{0,\ldots,P'-1\}$. Then,
the weighting vector corresponding to the learned dictionary is given by
$\hat{\vecw} =  \tp{(\hat{w}_0,\ldots,\hat{w}_{P'-1})}$ with
\begin{equation*}
     \hat{w}_{p} = \frac{\|\matM_{\text{R}}\hat{\vecg}_{p}\|_{2}^2}{\|{\hat{\vecg}_{p}}\|_{2}^2}.\vspace*{0mm}
 \end{equation*}
 
As in the original setting \cite{mokry2020audio}, we can also solve \eqref{eq:mod weighted l1 uncons} using the CP proximal algorithm. 
The learned weighted $\ell_1$-minimization algorithm for inpainting (CP-LEARNED) is summarized in Alg. \ref{CP-LEARNED}. 
\begin{algorithm}[t]\label{CP-LEARNED}
	\DontPrintSemicolon
	\KwInput{$\vecx$,
	$\matM_{\text{R}}$, $\hat{\matD}_{\text{block}}$, $\tilde{\matG}_{\text{A}}$, $\tilde{\matG}_{\text{S}}$,  $ \vecw_{\text{L}}$, 
	$\sigma$, $\tau$}
	\KwOutput{$\hat{\vecy}$}
	\SetKwBlock{repeat}{repeat}{end repeat}
	choose $\tau$, $\sigma > 0$  satisfying $\tau \sigma \|\tilde{\matG}_{\text{S}}\herm{\hat{\matD}_{\text{block}}} \| \leq 1$\\
	choose primal variable $\vecp^{(0)} \in \complexset^L$ and dual variable $\vecq^{(0)} \in \complexset^{N M'}$ arbitrarily\\
	set output variable $\vecz^{(0)}=\vecp^{(0)}$\\
	set iteration counter $i= 0$\\
    \repeat{
    $\vecq^{(i+1)} = \text{clip}_{\vecw_{\text{L}}}\left(\vecq^{(i)}+\sigma\cdot \hat{\matD}_{\text{block}}\tilde{\matG}_{\text{A}}\vecz^{(i)}\right)$\\
    $\vecp^{(i+1)} = \text{proj}_{\setS_{\vecx}}\left( \vecp^{(i)}-\tau\cdot \tilde{\matG}_{\text{S}} \herm{\hat{\matD}_{\text{block}}}\vecq^{(i+1)}
    \right)$\\
	${\vecz}^{(i+1)} =  2\vecp^{(i+1)}-\vecp^{(i)}$\\
	$i\gets i+1$}
	\textbf{until} \textit{stopping criterion met} \\
	\textbf{return} $\hat{\vecy} = \text{proj}_{\setS_{\vecx}}\left({\vecz}^{(i+1)} \right)$
	\caption{Learned weighted $\ell_1$-minimization (CP-LEARNED)}
\end{algorithm}
%

Note that we define   $\text{clip}_{\vecw_{\text{L}}}\left(\vecz\right)$, $\text{proj}_{\setS_{\vecx}}\left( \vecz\right)$, and stopping criterion essentially in the same way as in \cite{mokry2020audio}, 
\begin{align*}
  \text{clip}_{\vecw_{\text{L}}}\left(\vecz\right)&  = \vecz-\text{arg}(\vecz)\odot \text{max}(\lvert \vecz \rvert -\vecw_{\text{L}},\veczero),\\ \text{proj}_{\setS_{\vecx}}\left( \vecz\right) &  = \Re\left( (\matI-\matM_{\text{R}})\vecz + \matM_{\text{R}}\vecx\right),
\end{align*}
by, additionally, taking into account that $\setS_{\vecx}$ is real-valued.
The stopping criterion is
\begin{align*}
  {\vecnorm{\tilde{\matG}_{\text{A}}\vecz^{(i)} - \tilde{\matG}_{\text{A}}\vecz^{(i-1)}}_2^2} \leq  \epsilon {\vecnorm{\tilde{\matG}_{\text{A}}\vecz^{(i-1)}}_2^2},\\[-7mm]
\end{align*}
where $\epsilon > 0$ is the tolerance.

\section{Simulation Results and Discussion}
In this section, we compare the audio inpainting performance of our learned $\ell_1$-minimization technique, CP-LEARNED, with the original method, CP, presented in \cite{mokry2020audio}. We also compare our approach with other audio inpainting techniques such as A-SPAIN-MOD \cite{tauboeck2020audio}, A-SPAIN-LEARNED \cite{tauboeck2020audio}, A-SPAIN \cite{mokry2019introducing} and JANSSEN \cite{AR_audio86}. `A-' stands for analysis variant. Note that A-SPAIN uses a frame-wise DFT dictionary (redundancy $4$) whereas A-SPAIN-LEARNED and A-SPAIN-MOD use a Gabor dictionary (redundancy $4$). JANSSEN uses auto-regressive model of order $p=\min(3H+2,w_{\vecg}/3)$,  where $H$ denotes the number of missing samples within the current frame (window), $w_{\vecg}$ is the window length, and the number of iterations was set to $50$. In order to allow for a valid and comparable assessment, we consider essentially the same setup as in \cite{tauboeck2020audio} and \cite{mokry2020audio}. 

As a means to compare the performance, we use the signal-to-distortion ratio (SDR) \cite{adler12} defined as, 
\begin{align*}
  &\text{SDR}(\vecx_{\text{orig}},\vecx_{\text{inp}}) = 10 \, \log_{10} \frac{\vecnorm{\vecx_{\text{orig}}}_2^2}{\vecnorm{\vecx_{\text{orig}} - \vecx_{\text{inp}}}_2^2} \quad \text{[dB]},\\[-5mm]
\end{align*}
where $\vecx_{\text{orig}}$ and $\vecx_{\text{inp}}$ denote original and inpainted signal within the gaps, respectively. Note that higher SDR values imply better signal restoration. 

For our experiments, we used a collection of ten music recordings chosen from the EBU SQAM dataset \cite{EBU}. All chosen signals were sampled at
44.1 kHz and possessed different levels of sparsity with respect to
the original Gabor dictionary. In each test instance, the input was a signal with 5
gaps at random positions. The lengths of these gaps ranged from
5 ms up to 85 ms. For fixed lengths, the results over all ten signals
containing the 5 gaps were averaged.

Furthermore, we also calculated the PEMO-Q values \cite{huber2006pemo}, which
utilize a human auditory system model to assess the quality of the restored signal. Thus, the PEMO-Q criterion is closer to the subjective evaluation than the SDR. PEMO-Q consists of two quantities: \emph{objective difference grade} (ODG)  and \emph{perceptual similarity measure} (PSM). For simplicity, we restrict to ODG in this paper.
ODG can be interpreted as the degree of perceptual similarity of $\vecx_{\text{orig}}$ and $\vecx_{\text{inp}}$. Basically, ODG quantifies the perceptual impact of audio artifacts
in the restored signal and its value ranges from $-4$ (very annoying)
up to $0$ (imperceptible). 

Throughout our experiments,  we used the fast implementation of Gabor transforms available in the LTFAT toolbox \cite{LTFAT1,ltfatnote030}, and adopted its time-frequency conventions. The Gabor parameters used in our experiment are summarized in Table \ref{tab:gabor parameters}.
\begin{table}[t]   
\begin{center}
    \begin{tabular}{|c|c|}
        \hline
        \begin{tabular}[c]{@{}l@{}}\\[-2mm] Gabor Parameters \\[1mm] \end{tabular} & \begin{tabular}[c]{@{}l@{}}\\[-2mm] Used Value\\[1mm] \end{tabular} \\ \hline \hline
        window $\vecg$ & \emph{Hann} window\\
        window length $w_{\vecg}$ & 2800 samples ($\sim$64 ms) \\
        hop size $a$ & 700 samples \\
        number of modulations $M$ & 2800 \\ \hline
    \end{tabular}
\end{center}
	\caption{Gabor Parameters}\label{tab:gabor parameters}
\end{table}
The dictionary learning procedure proposed, see \cite{tauboeck2020audio} for algorithmic details, is run for at most $\text{iter}_{\max} = 20$ iterations with off-diagonal parameter $d=1$,
and the remaining parameters were set to  
$\rho_{\text{start}}=1$ and $\varepsilon=2^{-20}$. Finally, all SPAIN variants used the iteration parameters $s=t=1$ \cite{tauboeck2020audio}.

\begin{figure*}[t]
    \centering
\hspace*{0mm}    {\includegraphics[scale = 0.7]{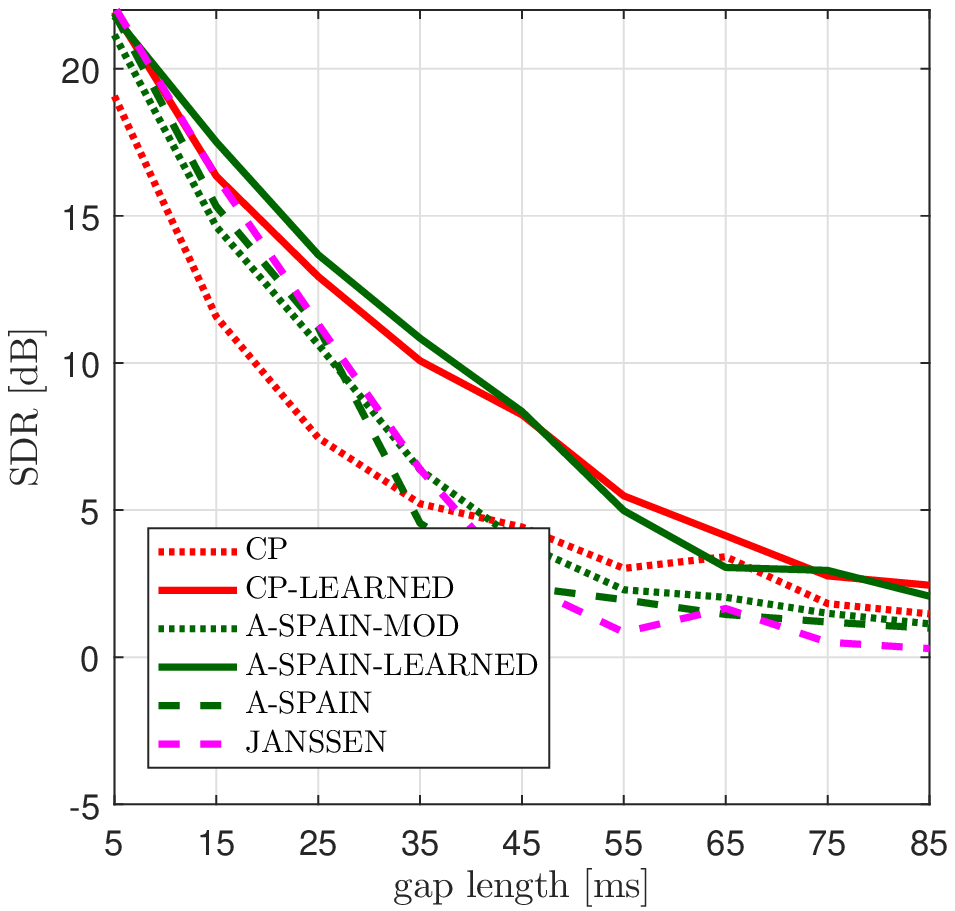}} \label{fig: SDR} 
\hspace*{10mm}    {\includegraphics[scale = 0.7]{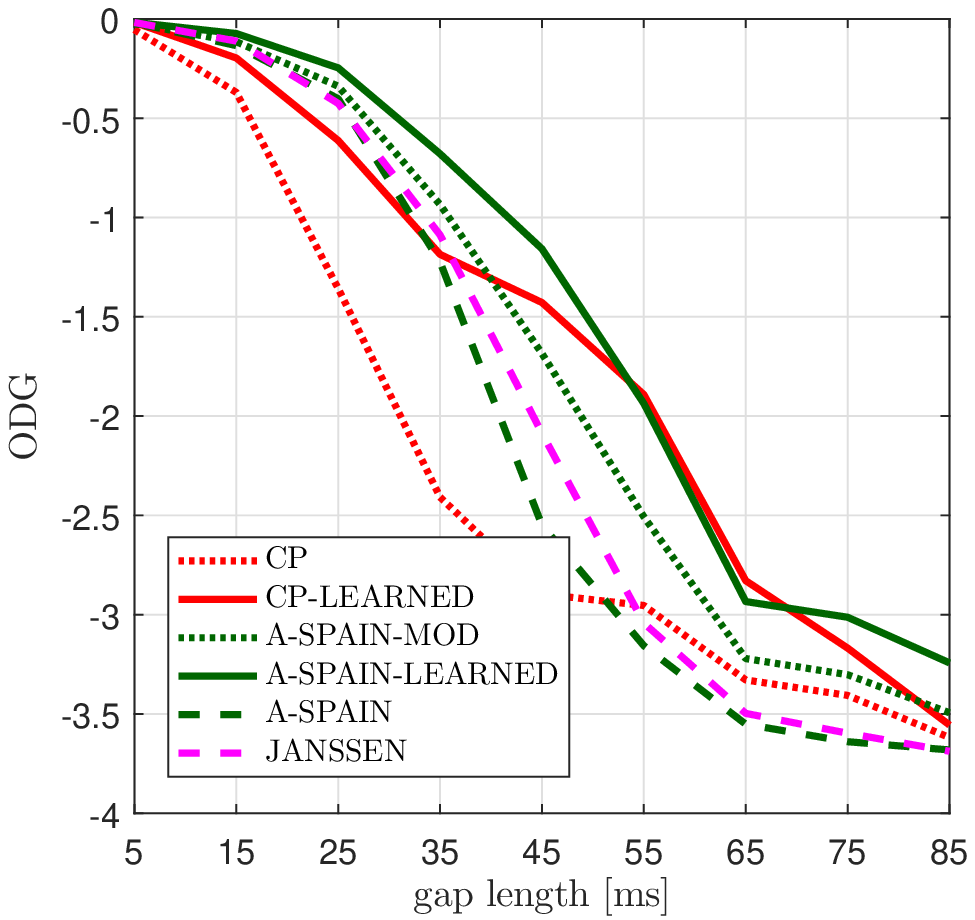}}\label{fig: ODG} 
\put(-355,-8){\small(a)} \put(-107,-8){\small(b)}  \vspace*{2mm}
       \caption{Performance comparison of different audio inpainting algorithms in terms of (a) SDR and (b) ODG for varying gap lengths.}
    \label{fig:perf}
     \vspace*{3mm}
\end{figure*}

Fig.\ \ref{fig:perf}(a) shows the inpainting performance in terms of SDR of all tested algorithms. We can observe that CP-LEARNED performs significantly better than CP, i.e., the original weighted $\ell_1$-minimization method. Even more, it beats all other methods except A-SPAIN-LEARNED, which is slightly superior for gap lengths below 45~ms. Remarkably, CP-LEARNED even outperforms the otherwise best performing A-SPAIN-LEARNED algorithm for longer gaps ($\geq$ 45~ms) although the observed improvements seem to be not very pronounced.

Fig.\ \ref{fig:perf}(b) shows the overall inpainting performance in terms of ODG. For smaller gap lengths ($\leq$ 35~ms), the performance of CP-LEARNED is superior to CP and a bit worse compared to other competing methods, whereas for longer gap lengths ($\geq$ 45~ms), CP-LEARNED's performance relative to the other methods improves, exceeding all the other methods except A-SPAIN-LEARNED for gap lengths above 65~ms.

\section{Conclusion}
\label{sec:prior}
We presented an approach for audio inpainting that combines weighted $\ell_1$-minimization with a dictionary learning framework in order to mitigate the energy drop phenomenon within the gap in the restored signal. This approach learns a sparsifying dictionary from the Gabor coefficients obtained from the reliable parts near the gap ultimately resulting in increased sparsity of the analysis coefficients. Additionally, the learned dictionary is also used for computing the weights. Moreover, by exploiting the real-valuedness of audio signals, we only used $M'=\lfloor M/2 \rfloor +1$ coefficients instead of the full $M$ coefficients for the inpainting task unlike in the original setting. This is noteworthy because utilizing half of the coefficients implies reduced computational effort. Our experimental results demonstrated that the proposed method, i.e, weighted $\ell_1$-minimization combined with dictionary learning, yields large performance gains and
significantly outperforms its original counterpart for all gap lengths. Finally, comparison with A-SPAIN-LEARNED demonstrated that CP-LEARNED has comparable (in some instances even better, in other slightly worse) performance as (than) the current state-of-the-art inpainting algorithm.

\bibliographystyle{IEEEtran}
\bibliography{references,MSPA_GT,RFP}

\end{document}

%% file: math-macros.tex
\usepackage{amsmath}
\usepackage{amssymb}
\usepackage{amsfonts}
\usepackage{amsthm}
\usepackage{mathrsfs}
\usepackage{xspace}
\usepackage{bm}
\usepackage{fancyref}
\usepackage{textcomp}
\usepackage[Symbol]{upgreek}

\newcommand{\safemath}[2]{\newcommand{#1}{\ensuremath{#2}\xspace}}

\newcommand{\ssa}{\mathsf{a}}
\newcommand{\ssb}{\mathsf{b}}
\newcommand{\ssc}{\mathsf{c}}
\newcommand{\ssd}{\mathsf{d}}
\newcommand{\sse}{\mathsf{e}}
\newcommand{\ssf}{\mathsf{f}}
\newcommand{\ssg}{\mathsf{g}}
\newcommand{\ssh}{\mathsf{h}}
\newcommand{\ssi}{\mathsf{i}}
\newcommand{\ssj}{\mathsf{j}}
\newcommand{\ssk}{\mathsf{k}}
\newcommand{\ssl}{\mathsf{l}}
\newcommand{\ssm}{\mathsf{m}}
\newcommand{\ssn}{\mathsf{n}}
\newcommand{\sso}{\mathsf{o}}
\newcommand{\ssp}{\mathsf{p}}
\newcommand{\ssq}{\mathsf{q}}
\newcommand{\ssr}{\mathsf{r}}
\newcommand{\sss}{\mathsf{s}}
\newcommand{\sst}{\mathsf{t}}
\newcommand{\ssu}{\mathsf{u}}
\newcommand{\ssv}{\mathsf{v}}
\newcommand{\ssw}{\mathsf{w}}
\newcommand{\ssx}{\mathsf{x}}
\newcommand{\ssy}{\mathsf{y}}
\newcommand{\ssz}{\mathsf{z}}

\safemath{\bmsa}{\bm{\ssa}}
\safemath{\bmsb}{\bm{\ssb}}
\safemath{\bmsc}{\bm{\ssc}}
\safemath{\bmsd}{\bm{\ssd}}
\safemath{\bmse}{\bm{\sse}}
\safemath{\bmsf}{\bm{\ssf}}
\safemath{\bmsg}{\bm{\ssg}}
\safemath{\bmsh}{\bm{\ssh}}
\safemath{\bmsi}{\bm{\ssi}}
\safemath{\bmsj}{\bm{\ssj}}
\safemath{\bmsk}{\bm{\ssk}}
\safemath{\bmsl}{\bm{\ssl}}
\safemath{\bmsm}{\bm{\ssm}}
\safemath{\bmsn}{\bm{\ssn}}
\safemath{\bmso}{\bm{\sso}}
\safemath{\bmsp}{\bm{\ssp}}
\safemath{\bmsq}{\bm{\ssq}}
\safemath{\bmsr}{\bm{\ssr}}
\safemath{\bmss}{\bm{\sss}}
\safemath{\bmst}{\bm{\sst}}
\safemath{\bmsu}{\bm{\ssu}}
\safemath{\bmsv}{\bm{\ssv}}
\safemath{\bmsw}{\bm{\ssw}}
\safemath{\bmsx}{\bm{\ssx}}
\safemath{\bmsy}{\bm{\ssy}}
\safemath{\bmsz}{\bm{\ssz}}

\bmdefine{\bmualphad}{\upalpha}
\bmdefine{\bmubetad}{\upbeta}
\bmdefine{\bmuchid}{\upchi}
\bmdefine{\bmudeltad}{\updelta}
\bmdefine{\bmuepsilond}{\upepsilon}
\bmdefine{\bmuvarepsilond}{\upvarepsilon}
\bmdefine{\bmuetad}{\upeta}
\bmdefine{\bmugammad}{\upgamma}
\bmdefine{\bmuiotad}{\upiota}
\bmdefine{\bmukappad}{\upkappa}
\bmdefine{\bmulambdad}{\uplambda}
\bmdefine{\bmumud}{\upmu}
\bmdefine{\bmunud}{\upnu}
\bmdefine{\bmuomegad}{\upomega}
\bmdefine{\bmuphid}{\upphi}
\bmdefine{\bmuvarphid}{\upvarphi}
\bmdefine{\bmupid}{\uppi}
\bmdefine{\bmuvarpid}{\upvarpi}
\bmdefine{\bmupsid}{\uppsi}
\bmdefine{\bmurhod}{\uprho}
\bmdefine{\bmuvarrhod}{\upvarrho}
\bmdefine{\bmusigmad}{\upsigma}
\bmdefine{\bmuvarsigmad}{\upvarsigma}
\bmdefine{\bmutaud}{\uptau}
\bmdefine{\bmuthetad}{\uptheta}
\bmdefine{\bmuvarthetad}{\upvartheta}
\bmdefine{\bmuupsilond}{\upupsilon}
\bmdefine{\bmuxid}{\upxi}
\bmdefine{\bmuzetad}{\upzeta}

\safemath{\bmua}{\mathbf{a}}
\safemath{\bmub}{\mathbf{b}}
\safemath{\bmuc}{\mathbf{c}}
\safemath{\bmud}{\mathbf{d}}
\safemath{\bmue}{\mathbf{e}}
\safemath{\bmuf}{\mathbf{f}}
\safemath{\bmug}{\mathbf{g}}
\safemath{\bmuh}{\mathbf{h}}
\safemath{\bmui}{\mathbf{i}}
\safemath{\bmuj}{\mathbf{j}}
\safemath{\bmuk}{\mathbf{k}}
\safemath{\bmul}{\mathbf{l}}
\safemath{\bmum}{\mathbf{m}}
\safemath{\bmun}{\mathbf{n}}
\safemath{\bmuo}{\mathbf{o}}
\safemath{\bmup}{\mathbf{p}}
\safemath{\bmuq}{\mathbf{q}}
\safemath{\bmur}{\mathbf{r}}
\safemath{\bmus}{\mathbf{s}}
\safemath{\bmut}{\mathbf{t}}
\safemath{\bmuu}{\mathbf{u}}
\safemath{\bmuv}{\mathbf{v}}
\safemath{\bmuw}{\mathbf{w}}
\safemath{\bmux}{\mathbf{x}}
\safemath{\bmuy}{\mathbf{y}}
\safemath{\bmuz}{\mathbf{z}}

\safemath{\bmualpha}{\bmualphad}
\safemath{\bmubeta}{\bmubetad}
\safemath{\bmuchi}{\bumchid}
\safemath{\bmudelta}{\bmudeltad}
\safemath{\bmuepsilon}{\bmuepsilond}
\safemath{\bmuvarepsilon}{\bmuvarepsilond}
\safemath{\bmueta}{\bmuetad}
\safemath{\bmugamma}{\bmugammad}
\safemath{\bmuiota}{\bmuiotad}
\safemath{\bmukappa}{\bmukappad}
\safemath{\bmulambda}{\bmulambdad}
\safemath{\bmumu}{\bmumud}
\safemath{\bmunu}{\bmunud}
\safemath{\bmuomega}{\bmuomegad}
\safemath{\bmuphi}{\bmuphid}
\safemath{\bmuvarphi}{\bmuvarphid}
\safemath{\bmupi}{\bmupid}
\safemath{\bmuvarpi}{\bmuvarpid}
\safemath{\bmupsi}{\bmupsid}
\safemath{\bmurho}{\bmurhod}
\safemath{\bmuvarrho}{\bmuvarrhod}
\safemath{\bmusigma}{\bmusigmad}
\safemath{\bmuvarsigma}{\bmuvarsigmad}
\safemath{\bmutau}{\bmutaud}
\safemath{\bmutheta}{\bmuthetad}
\safemath{\bmuvartheta}{\bmuvarthetad}
\safemath{\bmuupsilon}{\bmuupsilond}
\safemath{\bmuxi}{\bmuxid}
\safemath{\bmuzeta}{\bmuzetad}

\bmdefine{\bmiad}{a}
\bmdefine{\bmibd}{b}
\bmdefine{\bmicd}{c}
\bmdefine{\bmidd}{d}
\bmdefine{\bmied}{e}
\bmdefine{\bmifd}{f}
\bmdefine{\bmigd}{g}
\bmdefine{\bmihd}{h}
\bmdefine{\bmiid}{i}
\bmdefine{\bmijd}{j}
\bmdefine{\bmikd}{k}
\bmdefine{\bmild}{l}
\bmdefine{\bmimd}{m}
\bmdefine{\bmind}{n}
\bmdefine{\bmiod}{o}
\bmdefine{\bmipd}{p}
\bmdefine{\bmiqd}{q}
\bmdefine{\bmird}{r}
\bmdefine{\bmisd}{s}
\bmdefine{\bmitd}{t}
\bmdefine{\bmiud}{u}
\bmdefine{\bmivd}{v}
\bmdefine{\bmiwd}{w}
\bmdefine{\bmixd}{x}
\bmdefine{\bmiyd}{y}
\bmdefine{\bmizd}{z}

\bmdefine{\bmialphad}{\alpha}
\bmdefine{\bmibetad}{\beta}
\bmdefine{\bmichid}{\chi}
\bmdefine{\bmideltad}{\delta}
\bmdefine{\bmiepsilond}{\epsilon}
\bmdefine{\bmivarepsilond}{\varepsilon}
\bmdefine{\bmietad}{\eta}
\bmdefine{\bmigammad}{\gamma}
\bmdefine{\bmiiotad}{\iota}
\bmdefine{\bmikappad}{\kappa}
\bmdefine{\bmivarkappad}{\varkappa}
\bmdefine{\bmilambdad}{\lambda}
\bmdefine{\bmimud}{\mu}
\bmdefine{\bminud}{\nu}
\bmdefine{\bmiomegad}{\omega}
\bmdefine{\bmiphid}{\phi}
\bmdefine{\bmivarphid}{\varphi}
\bmdefine{\bmipid}{\pi}
\bmdefine{\bmivarpid}{\varpi}
\bmdefine{\bmipsid}{\psi}
\bmdefine{\bmirhod}{\rho}
\bmdefine{\bmivarrhod}{\varrho}
\bmdefine{\bmisigmad}{\sigma}
\bmdefine{\bmivarsigmad}{\varsigma}
\bmdefine{\bmitaud}{\tau}
\bmdefine{\bmithetad}{\theta}
\bmdefine{\bmivarthetad}{\vartheta}
\bmdefine{\bmiupsilond}{\upsilon}
\bmdefine{\bmixid}{\xi}
\bmdefine{\bmizetad}{\zeta}

\safemath{\bmia}{\bmiad}
\safemath{\bmib}{\bmibd}
\safemath{\bmic}{\bmicd}
\safemath{\bmid}{\bmidd}
\safemath{\bmie}{\bmied}
\safemath{\bmif}{\bmifd}
\safemath{\bmig}{\bmigd}
\safemath{\bmih}{\bmihd}
\safemath{\bmii}{\bmiid}
\safemath{\bmij}{\bmijd}
\safemath{\bmik}{\bmikd}
\safemath{\bmil}{\bmild}
\safemath{\bmim}{\bmimd}
\safemath{\bmin}{\bmind}
\safemath{\bmio}{\bmiod}
\safemath{\bmip}{\bmipd}
\safemath{\bmiq}{\bmiqd}
\safemath{\bmir}{\bmird}
\safemath{\bmis}{\bmisd}
\safemath{\bmit}{\bmitd}
\safemath{\bmiu}{\bmiud}
\safemath{\bmiv}{\bmivd}
\safemath{\bmiw}{\bmiwd}
\safemath{\bmix}{\bmixd}
\safemath{\bmiy}{\bmiyd}
\safemath{\bmiz}{\bmizd}

\safemath{\bmialpha}{\bmialphad}
\safemath{\bmibeta}{\bmibetad}
\safemath{\bmichi}{\bmichid}
\safemath{\bmidelta}{\bmideltad}
\safemath{\bmiepsilon}{\bmiepsilond}
\safemath{\bmivarepsilon}{\bmivarepsilond}
\safemath{\bmieta}{\bmietad}
\safemath{\bmigamma}{\bmigammad}
\safemath{\bmiiota}{\bmiiotad}
\safemath{\bmikappa}{\bmikappad}
\safemath{\bmivarkappa}{\bmivarkappad}
\safemath{\bmilambda}{\bmilambdad}
\safemath{\bmimu}{\bmimud}
\safemath{\bminu}{\bminud}
\safemath{\bmiomega}{\bmiomegad}
\safemath{\bmiphi}{\bmiphid}
\safemath{\bmivarphi}{\bmivarphid}
\safemath{\bmipi}{\bmipid}
\safemath{\bmivarpi}{\bmivarpid}
\safemath{\bmipsi}{\bmipsid}
\safemath{\bmirho}{\bmirhod}
\safemath{\bmivarrho}{\bmivarrhod}
\safemath{\bmisigma}{\bmisigmad}
\safemath{\bmivarsigma}{\bmivarsigmad}
\safemath{\bmitau}{\bmitaud}
\safemath{\bmitheta}{\bmithetad}
\safemath{\bmivartheta}{\bmivarthetad}
\safemath{\bmiupsilon}{\bmiupsilond}
\safemath{\bmixi}{\bmixid}
\safemath{\bmizeta}{\bmizetad}

\bmdefine{\bmuDeltad}{\Updelta}
\bmdefine{\bmuGammad}{\Upgamma}
\bmdefine{\bmuLambdad}{\Uplambda}
\bmdefine{\bmuOmegad}{\Upomega}
\bmdefine{\bmuPhid}{\Upphi}
\bmdefine{\bmuPid}{\Uppi}
\bmdefine{\bmuPsid}{\Uppsi}
\bmdefine{\bmuSigmad}{\Upsigma}
\bmdefine{\bmuThetad}{\Uptheta}
\bmdefine{\bmuUpsilond}{\Upupsilon}
\bmdefine{\bmuXid}{\Upxi}

\safemath{\bmuA}{\mathbf{A}}
\safemath{\bmuB}{\mathbf{B}}
\safemath{\bmuC}{\mathbf{C}}
\safemath{\bmuD}{\mathbf{D}}
\safemath{\bmuE}{\mathbf{E}}
\safemath{\bmuF}{\mathbf{F}}
\safemath{\bmuG}{\mathbf{G}}
\safemath{\bmuH}{\mathbf{H}}
\safemath{\bmuI}{\mathbf{I}}
\safemath{\bmuJ}{\mathbf{J}}
\safemath{\bmuK}{\mathbf{K}}
\safemath{\bmuL}{\mathbf{L}}
\safemath{\bmuM}{\mathbf{M}}
\safemath{\bmuN}{\mathbf{N}}
\safemath{\bmuO}{\mathbf{O}}
\safemath{\bmuP}{\mathbf{P}}
\safemath{\bmuQ}{\mathbf{Q}}
\safemath{\bmuR}{\mathbf{R}}
\safemath{\bmuS}{\mathbf{S}}
\safemath{\bmuT}{\mathbf{T}}
\safemath{\bmuU}{\mathbf{U}}
\safemath{\bmuV}{\mathbf{V}}
\safemath{\bmuW}{\mathbf{W}}
\safemath{\bmuX}{\mathbf{X}}
\safemath{\bmuY}{\mathbf{Y}}
\safemath{\bmuZ}{\mathbf{Z}}

\safemath{\bmuZero}{\mathbf{0}}
\safemath{\bmuOne}{\mathbf{1}}

\safemath{\bmuDelta}{\bmuDeltad}
\safemath{\bmuGamma}{\bmuGammad}
\safemath{\bmuLambda}{\bmuLambdad}
\safemath{\bmuOmega}{\bmuOmegad}
\safemath{\bmuPhi}{\bmuPhid}
\safemath{\bmuPi}{\bmuPid}
\safemath{\bmuPsi}{\bmuPsid}
\safemath{\bmuSigma}{\bmuSigmad}
\safemath{\bmuTheta}{\bmuThetad}
\safemath{\bmuUpsilon}{\bmuUpsilond}
\safemath{\bmuXi}{\bmuXid}

\bmdefine{\bmiAd}{A}
\bmdefine{\bmiBd}{B}
\bmdefine{\bmiCd}{C}
\bmdefine{\bmiDd}{D}
\bmdefine{\bmiEd}{E}
\bmdefine{\bmiFd}{F}
\bmdefine{\bmiGd}{G}
\bmdefine{\bmiHd}{H}
\bmdefine{\bmiId}{I}
\bmdefine{\bmiJd}{J}
\bmdefine{\bmiKd}{K}
\bmdefine{\bmiLd}{L}
\bmdefine{\bmiMd}{M}
\bmdefine{\bmiOd}{N}
\bmdefine{\bmiPd}{O}
\bmdefine{\bmiQd}{P}
\bmdefine{\bmiRd}{R}
\bmdefine{\bmiSd}{S}
\bmdefine{\bmiTd}{T}
\bmdefine{\bmiUd}{U}
\bmdefine{\bmiVd}{V}
\bmdefine{\bmiWd}{W}
\bmdefine{\bmiXd}{X}
\bmdefine{\bmiYd}{Y}
\bmdefine{\bmiZd}{Z}

\bmdefine{\bmiDeltad}{\Delta}
\bmdefine{\bmiGammad}{\Gamma}
\bmdefine{\bmiLambdad}{\Lambda}
\bmdefine{\bmiOmegad}{\Omega}
\bmdefine{\bmiPhid}{\Phi}
\bmdefine{\bmiPid}{\Pi}
\bmdefine{\bmiPsid}{\Psi}
\bmdefine{\bmiSigmad}{\Sigma}
\bmdefine{\bmiThetad}{\Theta}
\bmdefine{\bmiUpsilond}{\Upsilon}
\bmdefine{\bmiXid}{\Xi}

\safemath{\bmiA}{\bmiAd}
\safemath{\bmiB}{\bmiBd}
\safemath{\bmiC}{\bmiCd}
\safemath{\bmiD}{\bmiDd}
\safemath{\bmiE}{\bmiEd}
\safemath{\bmiF}{\bmiFd}
\safemath{\bmiG}{\bmiGd}
\safemath{\bmiH}{\bmiHd}
\safemath{\bmiI}{\bmiId}
\safemath{\bmiJ}{\bmiJd}
\safemath{\bmiK}{\bmiKd}
\safemath{\bmiL}{\bmiLd}
\safemath{\bmiM}{\bmiMd}
\safemath{\bmiN}{\bmiNd}
\safemath{\bmiO}{\bmiOd}
\safemath{\bmiP}{\bmiPd}
\safemath{\bmiQ}{\bmiQd}
\safemath{\bmiR}{\bmiRd}
\safemath{\bmiS}{\bmiSd}
\safemath{\bmiT}{\bmiTd}
\safemath{\bmiU}{\bmiUd}
\safemath{\bmiV}{\bmiVd}
\safemath{\bmiW}{\bmiWd}
\safemath{\bmiX}{\bmiXd}
\safemath{\bmiY}{\bmiYd}
\safemath{\bmiZ}{\bmiZd}

\safemath{\bmiDelta}{\bmiDeltad}
\safemath{\bmiGamma}{\bmiGammad}
\safemath{\bmiLambda}{\bmiLambdad}
\safemath{\bmiOmega}{\bmiOmegad}
\safemath{\bmiPhi}{\bmiPhid}
\safemath{\bmiPi}{\bmiPid}
\safemath{\bmiPsi}{\bmiPsid}
\safemath{\bmiSigma}{\bmiSigmad}
\safemath{\bmiTheta}{\bmiThetad}
\safemath{\bmiUpsilon}{\bmiUpsilond}
\safemath{\bmiXi}{\bmiXid}

\safemath{\evA}{\mathcal{A}}
\safemath{\evB}{\mathcal{B}}
\safemath{\evC}{\mathcal{C}}
\safemath{\evD}{\mathcal{D}}
\safemath{\evE}{\mathcal{E}}
\safemath{\evF}{\mathcal{F}}
\safemath{\evG}{\mathcal{G}}
\safemath{\evH}{\mathcal{H}}
\safemath{\evI}{\mathcal{I}}
\safemath{\evJ}{\mathcal{J}}
\safemath{\evK}{\mathcal{K}}
\safemath{\evL}{\mathcal{L}}
\safemath{\evM}{\mathcal{M}}
\safemath{\evN}{\mathcal{N}}
\safemath{\evO}{\mathcal{O}}
\safemath{\evP}{\mathcal{P}}
\safemath{\evQ}{\mathcal{Q}}
\safemath{\evR}{\mathcal{R}}
\safemath{\evS}{\mathcal{S}}
\safemath{\evT}{\mathcal{T}}
\safemath{\evU}{\mathcal{U}}
\safemath{\evV}{\mathcal{V}}
\safemath{\evW}{\mathcal{W}}
\safemath{\evX}{\mathcal{X}}
\safemath{\evY}{\mathcal{Y}}
\safemath{\evZ}{\mathcal{Z}}

\safemath{\setA}{\mathcal{A}}
\safemath{\setB}{\mathcal{B}}
\safemath{\setC}{\mathcal{C}}
\safemath{\setD}{\mathcal{D}}
\safemath{\setE}{\mathcal{E}}
\safemath{\setF}{\mathcal{F}}
\safemath{\setG}{\mathcal{G}}
\safemath{\setH}{\mathcal{H}}
\safemath{\setI}{\mathcal{I}}
\safemath{\setJ}{\mathcal{J}}
\safemath{\setK}{\mathcal{K}}
\safemath{\setL}{\mathcal{L}}
\safemath{\setM}{\mathcal{M}}
\safemath{\setN}{\mathcal{N}}
\safemath{\setO}{\mathcal{O}}
\safemath{\setP}{\mathcal{P}}
\safemath{\setQ}{\mathcal{Q}}
\safemath{\setR}{\mathcal{R}}
\safemath{\setS}{\mathcal{S}}
\safemath{\setT}{\mathcal{T}}
\safemath{\setU}{\mathcal{U}}
\safemath{\setV}{\mathcal{V}}
\safemath{\setW}{\mathcal{W}}
\safemath{\setX}{\mathcal{X}}
\safemath{\setY}{\mathcal{Y}}
\safemath{\setZ}{\mathcal{Z}}
\safemath{\emptySet}{\varnothing}

\safemath{\colA}{\mathscr{A}}
\safemath{\colB}{\mathscr{B}}
\safemath{\colC}{\mathscr{C}}
\safemath{\colD}{\mathscr{D}}
\safemath{\colE}{\mathscr{E}}
\safemath{\colF}{\mathscr{F}}
\safemath{\colG}{\mathscr{G}}
\safemath{\colH}{\mathscr{H}}
\safemath{\colI}{\mathscr{I}}
\safemath{\colJ}{\mathscr{J}}
\safemath{\colK}{\mathscr{K}}
\safemath{\colL}{\mathscr{L}}
\safemath{\colM}{\mathscr{M}}
\safemath{\colN}{\mathscr{N}}
\safemath{\colO}{\mathscr{O}}
\safemath{\colP}{\mathscr{P}}
\safemath{\colQ}{\mathscr{Q}}
\safemath{\colR}{\mathscr{R}}
\safemath{\colS}{\mathscr{S}}
\safemath{\colT}{\mathscr{T}}
\safemath{\colU}{\mathscr{U}}
\safemath{\colV}{\mathscr{V}}
\safemath{\colW}{\mathscr{W}}
\safemath{\colX}{\mathscr{X}}
\safemath{\colY}{\mathscr{Y}}
\safemath{\colZ}{\mathscr{Z}}

\safemath{\opA}{\mathbb{A}}
\safemath{\opB}{\mathbb{B}}
\safemath{\opC}{\mathbb{C}}
\safemath{\opD}{\mathbb{D}}
\safemath{\opE}{\mathbb{E}}
\safemath{\opF}{\mathbb{F}}
\safemath{\opG}{\mathbb{G}}
\safemath{\opH}{\mathbb{H}}
\safemath{\opI}{\mathbb{I}}
\safemath{\opJ}{\mathbb{J}}
\safemath{\opK}{\mathbb{K}}
\safemath{\opL}{\mathbb{L}}
\safemath{\opM}{\mathbb{M}}
\safemath{\opN}{\mathbb{N}}
\safemath{\opO}{\mathbb{O}}
\safemath{\opP}{\mathbb{P}}
\safemath{\opQ}{\mathbb{Q}}
\safemath{\opR}{\mathbb{R}}
\safemath{\opS}{\mathbb{S}}
\safemath{\opT}{\mathbb{T}}
\safemath{\opU}{\mathbb{U}}
\safemath{\opV}{\mathbb{V}}
\safemath{\opW}{\mathbb{W}}
\safemath{\opX}{\mathbb{X}}
\safemath{\opY}{\mathbb{Y}}
\safemath{\opZ}{\mathbb{Z}}
\safemath{\opZero}{\mathbb{O}}
\safemath{\identityop}{\opI}

\safemath{\sca}{a}
\safemath{\scb}{b}
\safemath{\scc}{c}
\safemath{\scd}{d}
\safemath{\sce}{e}
\safemath{\scf}{f}
\safemath{\scg}{g}
\safemath{\sch}{h}
\safemath{\sci}{i}
\safemath{\scj}{j}
\safemath{\sck}{k}
\safemath{\scl}{l}
\safemath{\scm}{m}
\safemath{\scn}{n}
\safemath{\sco}{o}
\safemath{\scp}{p}
\safemath{\scq}{q}
\safemath{\scr}{r}
\safemath{\scs}{s}
\safemath{\sct}{t}
\safemath{\scu}{u}
\safemath{\scv}{v}
\safemath{\scw}{w}
\safemath{\scx}{x}
\safemath{\scy}{y}
\safemath{\scz}{z}

\safemath{\scA}{A}
\safemath{\scB}{B}
\safemath{\scC}{C}
\safemath{\scD}{D}
\safemath{\scE}{E}
\safemath{\scF}{F}
\safemath{\scG}{G}
\safemath{\scH}{H}
\safemath{\scI}{I}
\safemath{\scJ}{J}
\safemath{\scK}{K}
\safemath{\scL}{L}
\safemath{\scM}{M}
\safemath{\scN}{N}
\safemath{\scO}{O}
\safemath{\scP}{P}
\safemath{\scQ}{Q}
\safemath{\scR}{R}
\safemath{\scS}{S}
\safemath{\scT}{T}
\safemath{\scU}{U}
\safemath{\scV}{V}
\safemath{\scW}{W}
\safemath{\scX}{X}
\safemath{\scY}{Y}
\safemath{\scZ}{Z}

\safemath{\scalpha}{\alpha}
\safemath{\scbeta}{\beta}
\safemath{\scchi}{\chi}
\safemath{\scdelta}{\delta}
\safemath{\scepsilon}{\epsilon}
\safemath{\scvarepsilon}{\varepsilon}
\safemath{\sceta}{\eta}
\safemath{\scgamma}{\gamma}
\safemath{\sciota}{\iota}
\safemath{\sckappa}{\kappa}
\safemath{\scvarkappa}{\varkappa}
\safemath{\sclambda}{\lambda}
\safemath{\scmu}{\mu}
\safemath{\scnu}{\nu}
\safemath{\scomega}{\omega}
\safemath{\scphi}{\phi}
\safemath{\scvarphi}{\varphi}
\safemath{\scpi}{\pi}
\safemath{\scvarpi}{\varpi}
\safemath{\scpsi}{\psi}
\safemath{\scrho}{\rho}
\safemath{\scvarrho}{\varrho}
\safemath{\scsigma}{\sigma}
\safemath{\scvarsigma}{\varsigma}
\safemath{\sctau}{\tau}
\safemath{\sctheta}{\theta}
\safemath{\scvartheta}{\vartheta}
\safemath{\scupsilon}{\upsilon}
\safemath{\scxi}{\xi}
\safemath{\sczeta}{\zeta}

\safemath{\veca}{\mathrm{a}}
\safemath{\vecb}{\mathrm{b}}
\safemath{\vecc}{\mathrm{c}}
\safemath{\vecd}{\mathrm{d}}
\safemath{\vece}{\mathrm{e}}
\safemath{\vecf}{\mathrm{f}}
\safemath{\vecg}{\mathrm{g}}
\safemath{\vech}{\mathrm{h}}
\safemath{\veci}{\mathrm{i}}
\safemath{\vecj}{\mathrm{j}}
\safemath{\veck}{\mathrm{k}}
\safemath{\vecl}{\mathrm{l}}
\safemath{\vecm}{\mathrm{m}}
\safemath{\vecn}{\mathrm{n}}
\safemath{\veco}{\mathrm{o}}
\safemath{\vecp}{\mathrm{p}}
\safemath{\vecq}{\mathrm{q}}
\safemath{\vecr}{\mathrm{r}}
\safemath{\vecs}{\mathrm{s}}
\safemath{\vect}{\mathrm{t}}
\safemath{\vecu}{\mathrm{u}}
\safemath{\vecv}{\mathrm{v}}
\safemath{\vecw}{\mathrm{w}}
\safemath{\vecx}{\mathrm{x}}
\safemath{\vecy}{\mathrm{y}}
\safemath{\vecz}{\mathrm{z}}
\safemath{\veczero}{\mathrm{0}}
\safemath{\vecone}{\mathrm{1}}

\safemath{\vecalpha}{\upalpha}
\safemath{\vecbeta}{\upbeta}
\safemath{\vecchi}{\upchi}
\safemath{\vecdelta}{\updelta}
\safemath{\vecepsilon}{\upepsilon}
\safemath{\vecvarepsilon}{\upvarepsilon}
\safemath{\veceta}{\upeta}
\safemath{\vecgamma}{\upgamma}
\safemath{\veciota}{\upiota}
\safemath{\veckappa}{\upkappa}
\safemath{\veclambda}{\uplambda}
\safemath{\vecmu}{\text{\textmu}}
\safemath{\vecnu}{\upnu}
\safemath{\vecomega}{\upomega}
\safemath{\vecphi}{\upphi}
\safemath{\vecvarphi}{\upvarphi}
\safemath{\vecpi}{\uppi}
\safemath{\vecvarpi}{\upvarpi}
\safemath{\vecpsi}{\uppsi}
\safemath{\vecrho}{\uprho}
\safemath{\vecvarrho}{\upvarrho}
\safemath{\vecsigma}{\upsigma}
\safemath{\vecvarsigma}{\upvarsigma}
\safemath{\vectau}{\uptau}
\safemath{\vectheta}{\uptheta}
\safemath{\vecvartheta}{\upvartheta}
\safemath{\vecupsilon}{\upupsilon}
\safemath{\vecxi}{\upxi}
\safemath{\veczeta}{\upzeta}

\safemath{\vecac}{a}
\safemath{\vecbc}{b}
\safemath{\veccc}{c}
\safemath{\vecdc}{d}
\safemath{\vecec}{e}
\safemath{\vecfc}{f}
\safemath{\vecgc}{g}
\safemath{\vechc}{h}
\safemath{\vecic}{i}
\safemath{\vecjc}{j}
\safemath{\veckc}{k}
\safemath{\veclc}{l}
\safemath{\vecmc}{m}
\safemath{\vecnc}{n}
\safemath{\vecoc}{o}
\safemath{\vecpc}{p}
\safemath{\vecqc}{q}
\safemath{\vecrc}{r}
\safemath{\vecsc}{s}
\safemath{\vectc}{t}
\safemath{\vecuc}{u}
\safemath{\vecvc}{v}
\safemath{\vecwc}{w}
\safemath{\vecxc}{x}
\safemath{\vecyc}{y}
\safemath{\veczc}{z}

\safemath{\matA}{\mathrm{A}}
\safemath{\matB}{\mathrm{B}}
\safemath{\matC}{\mathrm{C}}
\safemath{\matD}{\mathrm{D}}
\safemath{\matE}{\mathrm{E}}
\safemath{\matF}{\mathrm{F}}
\safemath{\matG}{\mathrm{G}}
\safemath{\matH}{\mathrm{H}}
\safemath{\matI}{\mathrm{I}}
\safemath{\matJ}{\mathrm{J}}
\safemath{\matK}{\mathrm{K}}
\safemath{\matL}{\mathrm{L}}
\safemath{\matM}{\mathrm{M}}
\safemath{\matN}{\mathrm{N}}
\safemath{\matO}{\mathrm{O}}
\safemath{\matP}{\mathrm{P}}
\safemath{\matQ}{\mathrm{Q}}
\safemath{\matR}{\mathrm{R}}
\safemath{\matS}{\mathrm{S}}
\safemath{\matT}{\mathrm{T}}
\safemath{\matU}{\mathrm{U}}
\safemath{\matV}{\mathrm{V}}
\safemath{\matW}{\mathrm{W}}
\safemath{\matX}{\mathrm{X}}
\safemath{\matY}{\mathrm{Y}}
\safemath{\matZ}{\mathrm{Z}}
\safemath{\matzero}{\mathrm{0}}
\safemath{\matpi}{\mathrm{\pi}}

\safemath{\matDelta}{\Updelta}
\safemath{\matGamma}{\Upgammma}
\safemath{\matLambda}{\Uplambda}
\safemath{\matOmega}{\Upomega}
\safemath{\matPhi}{\Upphi}
\safemath{\matPi}{\Uppi}
\safemath{\matPsi}{\Uppsi}
\safemath{\matSigma}{\Upsigma}
\safemath{\matTheta}{\Uptheta}
\safemath{\matUpsilon}{\Upupsilon}
\safemath{\matXi}{\Upxi}

\safemath{\matidentity}{\matI}
\safemath{\vecunit}{\vece} 
\safemath{\matone}{\matO}

\safemath{\matAc}{a}
\safemath{\matBc}{b}
\safemath{\matCc}{c}
\safemath{\matDc}{d}
\safemath{\matEc}{e}
\safemath{\matFc}{f}
\safemath{\matGc}{g}
\safemath{\matHc}{h}
\safemath{\matIc}{i}
\safemath{\matJc}{j}
\safemath{\matKc}{k}
\safemath{\matLc}{l}
\safemath{\matMc}{m}
\safemath{\matNc}{n}
\safemath{\matOc}{o}
\safemath{\matPc}{p}
\safemath{\matQc}{q}
\safemath{\matRc}{r}
\safemath{\matSc}{s}
\safemath{\matTc}{t}
\safemath{\matUc}{u}
\safemath{\matVc}{v}
\safemath{\matWc}{w}
\safemath{\matXc}{x}
\safemath{\matYc}{y}
\safemath{\matZc}{z}

\safemath{\rnda}{\mathsf{a}}
\safemath{\rndb}{\mathsf{b}}
\safemath{\rndc}{\mathsf{c}}
\safemath{\rndd}{\mathsf{d}}
\safemath{\rnde}{\mathsf{e}}
\safemath{\rndf}{\mathsf{f}}
\safemath{\rndg}{\mathsf{g}}
\safemath{\rndh}{\mathsf{h}}
\safemath{\rndi}{\mathsf{i}}
\safemath{\rndj}{\mathsf{j}}
\safemath{\rndk}{\mathsf{k}}
\safemath{\rndl}{\mathsf{l}}
\safemath{\rndm}{\mathsf{m}}
\safemath{\rndn}{\mathsf{n}}
\safemath{\rndo}{\mathsf{o}}
\safemath{\rndp}{\mathsf{p}}
\safemath{\rndq}{\mathsf{q}}
\safemath{\rndr}{\mathsf{r}}
\safemath{\rnds}{\mathsf{s}}
\safemath{\rndt}{\mathsf{t}}
\safemath{\rndu}{\mathsf{u}}
\safemath{\rndv}{\mathsf{v}}
\safemath{\rndw}{\mathsf{w}}
\safemath{\rndx}{\mathsf{x}}
\safemath{\rndy}{\mathsf{y}}
\safemath{\rndz}{\mathsf{z}}

\safemath{\rndA}{\bmiA}
\safemath{\rndB}{\bmiB}
\safemath{\rndC}{\bmiC}
\safemath{\rndD}{\bmiD}
\safemath{\rndE}{\bmiE}
\safemath{\rndF}{\bmiF}
\safemath{\rndG}{\bmiG}
\safemath{\rndH}{\bmiH}
\safemath{\rndI}{\bmiI}
\safemath{\rndJ}{\bmiJ}
\safemath{\rndK}{\bmiK}
\safemath{\rndL}{\bmiL}
\safemath{\rndM}{\bmiM}
\safemath{\rndN}{\bmiN}
\safemath{\rndO}{\bmiO}
\safemath{\rndP}{\bmiP}
\safemath{\rndQ}{\bmiQ}
\safemath{\rndR}{\bmiR}
\safemath{\rndS}{\bmiS}
\safemath{\rndT}{\bmiT}
\safemath{\rndU}{\bmiU}
\safemath{\rndV}{\bmiV}
\safemath{\rndW}{\bmiW}
\safemath{\rndX}{\bmiX}
\safemath{\rndY}{\bmiY}
\safemath{\rndZ}{\bmiZ}

\safemath{\rndalpha}{\bmialpha}
\safemath{\rndbeta}{\bmibeta}
\safemath{\rndchi}{\bmichi}
\safemath{\rnddelta}{\bmidelta}
\safemath{\rndepsilon}{\bmiepsilon}
\safemath{\rndvarepsilon}{\bmivarepsilon}
\safemath{\rndeta}{\bmieta}
\safemath{\rndgamma}{\bmigamma}
\safemath{\rndiota}{\bmiiota}
\safemath{\rndkappa}{\bmikappa}
\safemath{\rndlambda}{\bmilambda}
\safemath{\rndmu}{\bmimu}
\safemath{\rndnu}{\bminu}
\safemath{\rndomega}{\bmiomega}
\safemath{\rndphi}{\bmiphi}
\safemath{\rndvarphi}{\bmivarphi}
\safemath{\rndpi}{\bmipi}
\safemath{\rndvarpi}{\bmivarpi}
\safemath{\rndpsi}{\bmipsi}
\safemath{\rndrho}{\bmirho}
\safemath{\rndvarrho}{\bmivarrho}
\safemath{\rndsigma}{\bmisigma}
\safemath{\rndvarsigma}{\bmivarsigma}
\safemath{\rndtau}{\bmitau}
\safemath{\rndtheta}{\bmitheta}
\safemath{\rndvartheta}{\bmivartheta}
\safemath{\rndupsilon}{\bmiupsilon}
\safemath{\rndxi}{\bmixi}
\safemath{\rndzeta}{\bmizeta}

\safemath{\rveca}{\mathbf{a}}
\safemath{\rvecb}{\mathbf{b}}
\safemath{\rvecc}{\mathbf{c}}
\safemath{\rvecd}{\mathbf{d}}
\safemath{\rvece}{\mathbf{e}}
\safemath{\rvecf}{\mathbf{f}}
\safemath{\rvecg}{\mathbf{g}}
\safemath{\rvech}{\mathbf{h}}
\safemath{\rveci}{\mathbf{i}}
\safemath{\rvecj}{\mathbf{j}}
\safemath{\rveck}{\mathbf{k}}
\safemath{\rvecl}{\mathbf{l}}
\safemath{\rvecm}{\mathbf{m}}
\safemath{\rvecn}{\mathbf{n}}
\safemath{\rveco}{\mathbf{o}}
\safemath{\rvecp}{\mathbf{p}}
\safemath{\rvecq}{\mathbf{q}}
\safemath{\rvecr}{\mathbf{r}}
\safemath{\rvecs}{\mathbf{s}}
\safemath{\rvect}{\mathbf{t}}
\safemath{\rvecu}{\mathbf{u}}
\safemath{\rvecv}{\mathbf{v}}
\safemath{\rvecw}{\mathbf{w}}
\safemath{\rvecx}{\mathbf{x}}
\safemath{\rvecy}{\mathbf{y}}
\safemath{\rvecz}{\mathbf{z}}

\safemath{\rvecalpha}{\bmualpha}
\safemath{\rvecbeta}{\bmubeta}
\safemath{\rvecchi}{\bmuchi}
\safemath{\rvecdelta}{\bmudelta}
\safemath{\rvecepsilon}{\bmuepsilon}
\safemath{\rvecvarepsilon}{\bmuvarepsilon}
\safemath{\rveceta}{\bmueta}
\safemath{\rvecgamma}{\bmugamma}
\safemath{\rveciota}{\bmuiota}
\safemath{\rveckappa}{\bmukappa}
\safemath{\rveclambda}{\bmulambda}
\safemath{\rvecmu}{\bmumu}
\safemath{\rvecnu}{\bmunu}
\safemath{\rvecomega}{\bmuomega}
\safemath{\rvecphi}{\bmuphi}
\safemath{\rvecvarphi}{\bmuvarphi}
\safemath{\rvecpi}{\bmupi}
\safemath{\rvecvarpi}{\bmuvarpi}
\safemath{\rvecpsi}{\bmupsi}
\safemath{\rvecrho}{\bmurho}
\safemath{\rvecvarrho}{\bmuvarrho}
\safemath{\rvecsigma}{\bmusigma}
\safemath{\rvecvarsigma}{\bmuvarsigma}
\safemath{\rvectau}{\bmutau}
\safemath{\rvectheta}{\bmutheta}
\safemath{\rvecvartheta}{\bmuvartheta}
\safemath{\rvecupsilon}{\bmuupsilon}
\safemath{\rvecxi}{\bmuxi}
\safemath{\rveczeta}{\bmuzeta}

\safemath{\rvecac}{\rnda}
\safemath{\rvecbc}{\rndb}
\safemath{\rveccc}{\rndc}
\safemath{\rvecdc}{\rndd}
\safemath{\rvecec}{\rnde}
\safemath{\rvecfc}{\rndf}
\safemath{\rvecgc}{\rndg}
\safemath{\rvechc}{\rndh}
\safemath{\rvecic}{\rndi}
\safemath{\rvecjc}{\rndj}
\safemath{\rveckc}{\rndk}
\safemath{\rveclc}{\rndl}
\safemath{\rvecmc}{\rndm}
\safemath{\rvecnc}{\rndn}
\safemath{\rvecoc}{\rndo}
\safemath{\rvecpc}{\rndp}
\safemath{\rvecqc}{\rndq}
\safemath{\rvecrc}{\rndr}
\safemath{\rvecsc}{\rnds}
\safemath{\rvectc}{\rndt}
\safemath{\rvecuc}{\rndu}
\safemath{\rvecvc}{\rndv}
\safemath{\rvecwc}{\rndw}
\safemath{\rvecxc}{\rndx}
\safemath{\rvecyc}{\rndy}
\safemath{\rveczc}{\rndz}

\safemath{\rmatA}{\mathbf{A}}
\safemath{\rmatB}{\mathbf{B}}
\safemath{\rmatC}{\mathbf{C}}
\safemath{\rmatD}{\mathbf{D}}
\safemath{\rmatE}{\mathbf{E}}
\safemath{\rmatF}{\mathbf{F}}
\safemath{\rmatG}{\mathbf{G}}
\safemath{\rmatH}{\mathbf{H}}
\safemath{\rmatI}{\mathbf{I}}
\safemath{\rmatJ}{\mathbf{J}}
\safemath{\rmatK}{\mathbf{K}}
\safemath{\rmatL}{\mathbf{L}}
\safemath{\rmatM}{\mathbf{M}}
\safemath{\rmatN}{\mathbf{N}}
\safemath{\rmatO}{\mathbf{O}}
\safemath{\rmatP}{\mathbf{P}}
\safemath{\rmatQ}{\mathbf{Q}}
\safemath{\rmatR}{\mathbf{R}}
\safemath{\rmatS}{\mathbf{S}}
\safemath{\rmatT}{\mathbf{T}}
\safemath{\rmatU}{\mathbf{U}}
\safemath{\rmatV}{\mathbf{V}}
\safemath{\rmatW}{\mathbf{W}}
\safemath{\rmatX}{\mathbf{X}}
\safemath{\rmatY}{\mathbf{Y}}
\safemath{\rmatZ}{\mathbf{Z}}
\safemath{\rmatDelta}{\bmuDelta}
\safemath{\rmatGamma}{\bmuGamma}
\safemath{\rmatLambda}{\bmuLambda}
\safemath{\rmatOmega}{\bmuOmega}
\safemath{\rmatPhi}{\bmuPhi}
\safemath{\rmatPi}{\bmuPi}
\safemath{\rmatPsi}{\bmuPsi}
\safemath{\rmatSigma}{\bmuSigma}
\safemath{\rmatTheta}{\bmuTheta}
\safemath{\rmatUpsilon}{\bmuUpsilon}
\safemath{\rmatXi}{\bmuXi}

\safemath{\rmatAc}{\rnda}
\safemath{\rmatBc}{\rndb}
\safemath{\rmatCc}{\rndc}
\safemath{\rmatDc}{\rndd}
\safemath{\rmatEc}{\rnde}
\safemath{\rmatFc}{\rndf}
\safemath{\rmatGc}{\rndg}
\safemath{\rmatHc}{\rndh}
\safemath{\rmatIc}{\rndi}
\safemath{\rmatJc}{\rndj}
\safemath{\rmatKc}{\rndk}
\safemath{\rmatLc}{\rndl}
\safemath{\rmatMc}{\rndm}
\safemath{\rmatNc}{\rndn}
\safemath{\rmatOc}{\rndo}
\safemath{\rmatPc}{\rndp}
\safemath{\rmatQc}{\rndq}
\safemath{\rmatRc}{\rndr}
\safemath{\rmatSc}{\rnds}
\safemath{\rmatTc}{\rndt}
\safemath{\rmatUc}{\rndu}
\safemath{\rmatVc}{\rndv}
\safemath{\rmatWc}{\rndw}
\safemath{\rmatXc}{\rndx}
\safemath{\rmatYc}{\rndy}
\safemath{\rmatZc}{\rndz}

\newenvironment{textbmatrix}{	\setlength{\arraycolsep}{2.5pt}%
								\big[\begin{matrix}}{\end{matrix}\big]%
								\raisebox{0.08ex}{\vphantom{M}}}

 \def\btm{\begin{textbmatrix}}
 \def\etm{\end{textbmatrix}}

\DeclareMathOperator{\adj}{adj}				
\DeclareMathOperator*{\argmin}{arg\;min}		

\safemath{\fun}{\scf}						

\safemath{\vrbl}{x}						
\safemath{\altvrbl}{y}						
\safemath{\aaltvrbl}{z}						

\safemath{\vvrbl}{\vecx}						
\safemath{\altvvrbl}{\vecy}						
\safemath{\aaltvvrbl}{\vecz}						

\safemath{\altfun}{\scg}
\safemath{\aaltfun}{\sch}
\safemath{\bel}{\sce}					
\safemath{\altbel}{\sce}					
\safemath{\frel}{g}					
\safemath{\altfrel}{g}					
\safemath{\dfrel}{\tilde{g}}					
\safemath{\altdfrel}{\tilde{g}}					

\safemath{\mat}{\matA}						
\safemath{\matc}{\matAc}						
\safemath{\altmat}{\matB}						
\safemath{\altmatc}{\matBc}						

\safemath{\vectr}{\vecu}						
\safemath{\vectrc}{\vecuc}						
\safemath{\altvectr}{\vecv}						
\safemath{\aaltvectr}{\vect}						
\safemath{\altvectrc}{\vecvc}						
\safemath{\genvar}{u}						
\safemath{\altgenvar}{v}						

\safemath{\rvectr}{\rvecu}						
\safemath{\rvectrc}{\rvecuc}						
\safemath{\raltvectr}{\rvecv}						
\safemath{\raaltvectr}{\rvect}						
\safemath{\raltvectrc}{\rvecvc}						
\safemath{\rgenvar}{\rndu}						
\safemath{\raltgenvar}{\rndv}						

\newcommand{\nullspace}{\setN}	 			

\DeclareMathOperator{\card}{card}			

\newcommand{\vecnorm}[1]{\lVert#1\rVert}		
\newcommand{\conj}[1]{\ensuremath{#1^{*}}} 	
\newcommand{\tp}[1]{\ensuremath{#1^{\mathsf{T}}}} 		

\newcommand{\herm}[1]{\ensuremath{#1^{\mathsf{H}}}} 	
\newcommand{\inv}[1]{\ensuremath{#1^{-1}}} 	

\safemath{\dirac}{\delta}					
\safemath{\diracp}{\dirac(\time)}			
\safemath{\krond}{\dirac}					
\safemath{\indfun}{I}						
\safemath{\stepfun}{u}						

\safemath{\upto}{\uparrow}
\safemath{\downto}{\downarrow}
\safemath{\iu}{\mathrm{i}}							
\safemath{\maj}{\succ}
\newcommand{\dftmat}[1]{\matF_{#1}}			
\safemath{\mdft}{\dftmat{}}					
\safemath{\runity}{\beta}					
\safemath{\eval}{\lambda}					

\safemath{\veval}{\veclambda}				
\safemath{\littleo}{\sco}					

\let\im\undefined
\safemath{\re}{\Re}				
\safemath{\im}{\Im}				

\safemath{\euclidspace}{\complexset}			
\safemath{\confunspace}{\setC}				
\newcommand{\banachseqspace}[1]{l^{#1}}		
\safemath{\hilseqspace}{\banachseqspace{2}}	
\newcommand{\banachfunspace}[1]{\setL^{#1}}	
\safemath{\hilfunspace}{\banachfunspace{2}}	
\safemath{\hilfunspacep}{\hilfunspace(\complexset)}	
\safemath{\schwarzspace}{\setS}				

\newcommand{\hadj}[1]{#1^{\star}}			

\safemath{\SNR}{\text{\sc snr}} 				
\safemath{\SINR}{\text{\sc sinr}} 				
\safemath{\No}{N_0}							
\safemath{\Es}{E_s}							
\safemath{\Eb}{E_b}							
\safemath{\EbNo}{\frac{\Eb}{\No}}
\safemath{\EsNo}{\frac{\Es}{\No}}
\safemath{\NoVar}{\variance}                 

\let\time\undefined
\safemath{\time}{\sct}						
\safemath{\dtime}{\sck}						
\safemath{\delay}{\sctau}					
\safemath{\ddelay}{\scl}						
\safemath{\doppler}{\scnu}					
\safemath{\ddoppler}{\scm}					
\safemath{\freq}{\scf}						
\safemath{\dfreq}{\scn}						
\safemath{\Dtime}{\Delta\time}
\safemath{\Dfreq}{\Delta\freq}
\safemath{\Ddtime}{\dtime}
\safemath{\Ddfreq}{\dfreq}
\safemath{\bandwidth}{\scB}
\safemath{\maxdoppler}{\doppler_{0}}			
\safemath{\maxdelay}{\delay_{0}}				
\safemath{\spread}{\Delta_{\CHop}}			
\DeclareMathOperator{\CHop}{\ensuremath{\opH}} 
\safemath{\kernel}{\rndk_{\CHop}}			
\safemath{\kernelp}{\kernel(\time,\time')}	
\safemath{\tvir}{\rndh_{\CHop}}				
\safemath{\tvirp}{\tvir(\time,\delay)}		
\safemath{\tvirc}{\conj{\rndh}_{\CHop}}		
\safemath{\tvtf}{\rndl_{\CHop}}				
\safemath{\tvtfp}{\tvtf(\time,\freq)}			
\safemath{\tvtfc}{\conj{\rndl}_{\CHop}}		
\safemath{\spf}{\rnds_{\CHop}}				
\safemath{\spfp}{\spf(\doppler,\delay)}		
\safemath{\spfc}{\conj{\rnds}_{\CHop}}		
\safemath{\bff}{\rndb_{\CHop}}				
\safemath{\bffp}{\bff(\doppler,\freq)}		

\safemath{\irc}{\scr_{\rndh}}				
\safemath{\tfc}{\scr_{\rndl}}				
\safemath{\spc}{\scr_{\rnds}}				
\safemath{\bfc}{\scr_{\rndb}}				

\safemath{\scaf}{\scc_{\rnds}}				
\safemath{\scafp}{\scaf(\doppler,\delay)}		
\safemath{\ccf}{\scc_{\rndl}}				
\safemath{\ccfp}{\ccf(\Dtime,\Dfreq)}			
\safemath{\cic}{\scc_{\rndh}}				
\safemath{\cicp}{\cic(\Dtime,\delay)}			

\safemath{\mi}{I}							
\safemath{\capacity}{C}					

\DeclareMathOperator{\Prob}{\opP}		

\safemath{\normal}{\mathcal{N}}			
\safemath{\jpg}{\mathcal{CN}}			
\safemath{\uniform}{\mathcal{U}}				
\safemath{\mchain}{\leftrightarrow}		

\safemath{\dB}{\,\mathrm{dB}}
\safemath{\dBm}{\,\mathrm{dBm}}
\safemath{\Hz}{\,\mathrm{Hz}}
\safemath{\kHz}{\,\mathrm{kHz}}
\safemath{\MHz}{\,\mathrm{MHz}}
\safemath{\GHz}{\,\mathrm{GHz}}
\safemath{\s}{\,\mathrm{s}}
\safemath{\ms}{\,\mathrm{ms}}
\safemath{\mus}{\,\mathrm{\text{\textmu}s}}
\safemath{\ns}{\,\mathrm{ns}}
\safemath{\ps}{\,\mathrm{ps}}
\safemath{\meter}{\,\mathrm{m}}
\safemath{\mm}{\,\mathrm{mm}}
\safemath{\cm}{\,\mathrm{cm}}
\safemath{\m}{\,\mathrm{m}}
\safemath{\W}{\,\mathrm{W}}
\safemath{\mW}{\, \mathrm{mW}}
\safemath{\J}{\,\mathrm{J}}
\safemath{\K}{\,\mathrm{K}}
\safemath{\bit}{\,\mathrm{bit}}
\safemath{\nat}{\,\mathrm{nat}}

\safemath{\define}{\triangleq}					

\safemath{\equivalent}{\sim}
\safemath{\distas}{\sim}					
\safemath{\sdiff}{\Delta}				
\safemath{\setdiff}{\setminus}				

\safemath{\reals}{\mathbb R}
\safemath{\positivereals}{\reals^{+}}
\safemath{\integers}{\mathbb Z}
\safemath{\posint}{\integers^{+}}
\safemath{\naturals}{\mathbb N}
\safemath{\posnaturals}{\naturals^{+}}
\safemath{\complexset}{\mathbb C}
\safemath{\rationals}{\mathbb Q}

\safemath{\iSet}{\setI}
\safemath{\rel}{\bowtie}					
\safemath{\eqrel}{\sim}					
\safemath{\rlord}{\leq}					
\safemath{\slord}{<}						
\safemath{\rpord}{\preceq}				
\safemath{\rrpord}{\succeq}				
\safemath{\spord}{\prec}					
\safemath{\sig}{\sigma}					
\safemath{\metric}{d}					

\safemath{\setfun}{\Phi}					
\safemath{\measure}{\mu}					
\safemath{\altmeasure}{\lambda}					
\newcommand{\outerm}[1]{#1^{\star}}		
\newcommand{\innerm}[1]{#1_{\star}}		
\safemath{\omeasure}{\outerm{\measure}}		
\safemath{\imeasure}{\innerm{\measure}}		
\safemath{\aecol}{\colS^{\star}_{\measure}} 
\safemath{\emeasure}{\bar{\measure}_{0}}	
\safemath{\rmeasure}{\tilde{\measure}}	
\safemath{\bmeasure}{\measure_{0}}		
\safemath{\glength}{\measure_{\altfun}}	
\safemath{\lebmea}{\lambda}				
\safemath{\blebmea}{\lebmea_{0}}			
\safemath{\sfun}{s}						
\safemath{\absintspace}{\colL^{1}}		
\safemath{\sqintspace}{\colL^{2}}		
\safemath{\abssumspace}{l^{1}}		
\safemath{\sqsumspace}{l^{2}}		

\safemath{\sfield}{\setF}				
\safemath{\vectors}{\setV}				
\safemath{\vecspace}{(\vectors,\sfield)}	
\safemath{\linspace}{\setV}				
\safemath{\clinspace}{(\linspace,\sfield)} 
\safemath{\nspace}{\setU}				
\safemath{\metspace}{\setM}				
\safemath{\bspace}{\setB}				
\safemath{\ipspace}{\setG}				
\safemath{\hilspace}{\setH}				
\safemath{\blospace}{\setG}				
\safemath{\lop}{\opT}					
\safemath{\altlop}{\opS}					
\safemath{\nullsp}{\nullspace(\lop)}		
\safemath{\lfun}{l}						
\safemath{\altlfun}{g}					
\newcommand{\dual}[1]{#1^{'}}			
\safemath{\dsum}{\oplus}					
\safemath{\funspace}{\colL}				
\renewcommand{\adj}[1]{#1^{\times}}		
\safemath{\adjlop}{\adj{\lop}}			
\safemath{\hadjlop}{\hadj{\lop}}			
\safemath{\tow}{\xrightarrow{w}}			
\safemath{\tows}{\xrightarrow{w^{*}}}		
\safemath{\cparam}{\lambda}
\safemath{\lopl}{\lop_{\cparam}}		
\safemath{\iop}{\opI}					
\safemath{\resolop}{\opR}				
\safemath{\resolvent}{\resolop_{\cparam}(\lop)}	
\safemath{\reset}{\setQ}
\safemath{\spectrum}{\setS}
\safemath{\resolset}{\reset(\lop)}		
\safemath{\lopspec}{\spectrum(\lop)}		
\safemath{\pspec}{\spectrum_{p}(\lop)}	
\safemath{\cspec}{\spectrum_{c}(\lop)}	
\safemath{\rspec}{\spectrum_{r}(\lop)}	
\safemath{\ev}{\cparam}					
\newcommand{\specrad}[1]{r_{#1}}			
\safemath{\lopsrad}{\specrad{\lop}}		
\safemath{\pop}{\opP}					

\safemath{\specfam}{\colE}				
\safemath{\specop}{\opE_{\cparam}}		
\safemath{\altspecop}{\opE_{\mu}}		
\safemath{\vmulti}{\vecone}				
\safemath{\unitaryop}{\opU}				
\safemath{\sval}{\sigma}					
\safemath{\corrcoef}{\rho}				
\safemath{\sangle}{\theta}				

\let\time\undefined
\safemath{\iset}{\setI}				
\safemath{\shift}{\nu}
\safemath{\scale}{\alpha}
\safemath{\time}{t}
\safemath{\specfreq}{\theta}	
\newcommand{\transopgen}[1]{\opT_{#1}} 
\safemath{\transop}{\transopgen{\delay}}
\newcommand{\modopgen}[1]{\opM_{#1}}	
\safemath{\modop}{\modopgen{\shift}}
\newcommand{\dilopgen}[1]{\opD_{#1}}	
\safemath{\dilop}{\dilopgen{\scale}}
\safemath{\fram}{\setF}				
\safemath{\dfram}{\dual{\fram}}		
\safemath{\ufb}{B}					
\safemath{\lfb}{A}					
\safemath{\sop}{\hadj{\aop}}				
\safemath{\aop}{\opT}			
\safemath{\fop}{\opS}				
\safemath{\daop}{\tilde\opT}			
\safemath{\dsop}{\hadj{\tilde\opT}}				

\safemath{\ifop}{\inv{\fop}}			
\safemath{\rifop}{\fop^{-1/2}}			
\safemath{\transeq}{\setT}			
\safemath{\nfun}{\Phi}				
\safemath{\funvec}{\vecf}			
\safemath{\altfunvec}{\vecg}

\safemath{\samplespace}{\Omega}
\safemath{\probspace}{(\samplespace,\sfield,\Prob)}	
\safemath{\ccoef}{\rho}			

\safemath{\infstate}{\vecpi}				
\safemath{\typset}{\setA_{\epsilon}^{(N)}}	
\safemath{\expequal}{\doteq}				
\safemath{\code}{C}						
\safemath{\dstringset}{\setD^{\star}}		
\safemath{\cwlength}{l}					
\safemath{\elength}{L}					
\safemath{\extension}{C^{\star}}			
\safemath{\approaches}{\rightarrow}		
\safemath{\evnt}{\setA}					
\safemath{\altevnt}{\setB}					
\safemath{\rv}{\rndx}					
\safemath{\altrv}{\rndy}					
\safemath{\complexrv}{\rndu}					
\safemath{\altcrv}{\rndv}				
\safemath{\rvec}{\rvecx}					
\safemath{\altrvec}{\rvecy}				
\safemath{\crvec}{\rvecu}				
\safemath{\altcrvec}{\rvecv}				
\safemath{\variance}{\sigma^{2}}			
\safemath{\map}{T}						
\safemath{\jacobian}{\matJ}					
\safemath{\wvec}{\rvecw}					
\safemath{\wrv}{\rndw}					
\safemath{\orthmat}{\matQ}				
\safemath{\evmat}{\matLambda}			
\safemath{\identity}{\matidentity}		
\safemath{\innovec}{\vecv}				
\safemath{\convas}{\xrightarrow{\text{a.s.}}}	
\safemath{\convr}{\xrightarrow{\text{r}}}	
\safemath{\convp}{\xrightarrow{\text{P}}}	
\safemath{\convd}{\xrightarrow{\text{D}}}	
\safemath{\ltis}{\opL}				
\safemath{\ir}{h}					
\safemath{\tf}{\MakeUppercase{\ir}}	

\newcommand*{\fancyrefparlabelprefix}{par}		
\newcommand*{\fancyrefremlabelprefix}{rem}		
\newcommand*{\fancyrefchalabelprefix}{cha}		
\newcommand*{\fancyrefapplabelprefix}{app}		
\newcommand*{\fancyrefthmlabelprefix}{thm}		
\newcommand*{\fancyreflemlabelprefix}{lem}		
\newcommand*{\fancyrefcorlabelprefix}{cor}		
\newcommand*{\fancyrefdeflabelprefix}{def}		
\newcommand*{\fancyrefproplabelprefix}{prop}		

\frefformat{vario}{\fancyrefparlabelprefix}{Part~#1}
\frefformat{vario}{\fancyrefremlabelprefix}{Remark~#1}
\frefformat{vario}{\fancyrefchalabelprefix}{Chapter~#1}
\frefformat{vario}{\fancyrefseclabelprefix}{Section~#1}
\frefformat{vario}{\fancyrefthmlabelprefix}{Theorem~#1}
\frefformat{vario}{\fancyreflemlabelprefix}{Lemma~#1}
\frefformat{vario}{\fancyrefcorlabelprefix}{Corollary~#1}
\frefformat{vario}{\fancyrefdeflabelprefix}{Definition~#1}
\frefformat{vario}{\fancyreffiglabelprefix}{Figure~#1}
\frefformat{vario}{\fancyrefapplabelprefix}{Appendix~#1}
\frefformat{vario}{\fancyrefeqlabelprefix}{(#1)}
\frefformat{vario}{\fancyrefproplabelprefix}{Property~#1}

\theoremstyle{plain}

\theoremstyle{remark}

